\documentclass[twocolumn,showpacs,preprintnumbers,amsmath,amssymb]{revtex4}
\usepackage{graphicx}
\usepackage{dcolumn}
\usepackage{bm}
\def\bea {\begin{eqnarray}}
\def\eea {\end{eqnarray}}

\def\be {\begin{equation}}
\def\ee {\end{equation}}

\newcommand{\Gm}{\Gamma}

\newcommand{\del}{\partial}

\newcommand{\F}{F_\pi}
\newcommand{\D}{{\cal D}}
\newcommand{\oK}{\overline{K}}

\begin{document}

\title{Dragging $D$ mesons by hot hadrons}

\author{Sabyasachi Ghosh, Santosh K Das, Sourav Sarkar and  Jan-e Alam}
\medskip
\affiliation{Variable Energy Cyclotron Centre, 1/AF, Bidhan Nagar , 
Kolkata - 700064}

\date{\today}
\begin{abstract}
We evaluate the drag and
diffusion coefficients of a hot hadronic medium consisting of
pions, nucleons, kaons and eta using open charm mesons as a probe.
The interaction of the probe with the  hadronic matter 
has been treated in the framework of effective field theory.
It is observed that the magnitude of both the transport coefficients  
are significant, indicating substantial amount of interaction of the
heavy mesons with the thermal hadronic system. The 
results may have noticeable impact on the  experimental 
observables like the suppression of  single electron spectra originating 
form the decays of heavy mesons in nuclear collisions at relativistic energies.   
\end{abstract}

\maketitle

Nuclear collisions at Relativistic Heavy Ion Collider (RHIC) and Large Hadron
Collider(LHC) energies are aimed at creating a phase where the bulk properties of the
matter are governed by a deconfined state of (light) quarks and gluons known as
Quark Gluon Plasma (QGP). The study of the transport properties of QGP 
is a field of great contemporary interest and the heavy flavors, namely, 
charm and bottom quarks, play a crucial role in such studies. 
The  weakly interacting picture of the QGP 
stems from the perception of 
asymptotic freedom of QCD at high temperatures and densities.
However, the experimental data  from RHIC, especially 
the measured elliptic flow
indicate that the matter produced in Au+Au collisions exhibit
properties which are more like a strongly interacting liquid than 
a weakly interacting gas. The magnitude of the transport coefficients 
can be used to understand the strength of the interaction within the QGP. 
For example, the shear viscosity or the internal friction of the fluid
symbolizes the ability 
to transfer momentum over a distance of about a mean free path.
Therefore, in a system where the constituents interact strongly 
the transfer of momentum is performed easily 
- resulting in lower values of $\eta$.
Consequently such a system may be characterized by a small value of  
$\eta/s$ where $s$ is the entropy density. 
On the other hand, for a weakly interacting system the momentum
transfer between the constituents become strenuous which gives rise
to a large $\eta$. 
The importance of viscosity also lies in the fact that
it damps out the variation in the velocity and makes the fluid flow
laminar. A very small viscosity (large Reynold number) may make the
flow turbulent.  A lower bound on the  value of $\eta/s$ has recently been 
found using AdS/CFT~\cite{KSS}.  

The interaction of heavy quarks with the QGP can be used to estimate 
the  value of the transport coefficients. This has recently been
performed by experimentally 
measuring  the nuclear suppression factor 
($R_{\mathrm AA}$)~\cite{raaexpt} and 
the elliptic flow ($v_2$)~\cite{v2expt}
for the single electron spectra originating from the semi-leptonic 
decays of the heavy mesons which are
produced from heavy quark fragmentation.  
Several theoretical 
attempts~\cite{hvh1,hvh2,ko,adil,gossiaux,wicks,das,hirano,alberico,ars,teaney,Das3} 
have been made to
explain $R_{\mathrm AA}$ and $v_2$, where the role of 
hadronic matter has been ignored. However, to make the
characterization of QGP reliable the role of the
hadronic phase should be taken into consideration 
and its contribution must be subtracted out from the observables.  
Although a large amount of work has been done 
on the diffusion of heavy quarks in QGP, the diffusion of
heavy mesons in hadronic matter has received much
less attention so far. Recently the diffusion coefficient 
of $D$ meson has been  evaluated  using heavy meson 
chiral perturbation theory~\cite{laine} and also by using the empirical 
elastic scattering amplitudes~\cite{MinHe} of $D$ mesons with thermal hadrons.


The present work addresses the relevance of the hadronic sector to 
some of these issues. The drag and diffusion coefficients for the hadronic 
phase have been evaluated and their importance to the
experimental observables in heavy ion collisions 
have been discussed.
In particular, we consider the interaction
of a $D$ meson with a thermal hadronic system 
composed of pions, nucleons, kaons and $\eta$ in a temperature 
domain relevant for heavy ion phenomenology. 
It is expected that the  relaxation time
for heavy mesons are larger than the corresponding quantities
for light hadrons.  
The abundance of the heavy mesons 
will be low for the temperature ($T$) range under consideration (T=100-180 MeV),
as a result they  do not decide the bulk properties of the matter.
The thermal production of charm mesons can be ignored
for the range of temperature mentioned above.
Therefore, the drag ($\gamma$) and diffusion ($B_0$)
coefficients of the heavy mesons
can be evaluated by using its elastic interaction
with the thermal hadrons.
For the process, $D(p) + h(q) \rightarrow D(p^\prime) + h(q^\prime)$ 
($h$ stands for pion, nucleon, kaon and eta),
the drag $\gamma$ can be calculated by using the following 
expression~\cite{BS}:
\begin{equation}
\gamma=p_iA_i/p^2
\end{equation}
where $A_i$ is given by 
\begin{eqnarray}
A_i=\frac{1}{2E_p} \int \frac{d^3q}{(2\pi)^3E_q×} \int \frac{d^3p^\prime}{(2\pi)^3E_p^\prime×}
\int \frac{d^3q^\prime}{(2\pi)^3E_q^\prime×}  \nonumber \\ \frac{1}{g_D} 
\sum  \overline{|M|^2} (2\pi)^4 \delta^4(p+q-p^\prime-q^\prime) \nonumber \\
{f}(q)[(p-p^\prime)_i] \equiv \langle \langle
(p-p^\prime)\rangle \rangle
\label{eq1}
\end{eqnarray}
$g_D$ being the statistical degeneracy of the $D$ meson propagating in the medium.
The above expression indicates that the drag coefficient is the
measure of the thermal average of the square of the invariant amplitude
$\overline{\mid M\mid^2}$ weighted by the momentum transfer, $p-p^\prime$. 
The factor $f(q)$ denotes the thermal phase space for the particle
in the medium. 

Similarly the diffusion coefficient $B_0$ can be defined as:
\begin{equation}
B_0=\frac{1}{4}\left[\langle \langle p\prime^2 \rangle \rangle -
\frac{\langle \langle (p.p\prime)^2 \rangle \rangle }{p^2}\right]
\label{diffusion}
\end{equation}

\begin{figure}[ht]
\begin{center}
\includegraphics[scale=0.5]{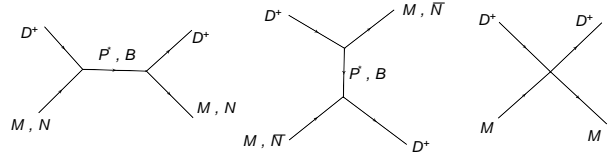}
\caption{Feynman diagrams for the scattering of $D$ mesons with
hadrons in the medium. Here $M$ stands for mesons (pion, kaon and eta) 
and $P^*$ and $B$ denote charmed vector mesons and baryonic resonances
respectively.
}
\label{fig0}
\end{center}
\end{figure}

With an appropriate choice of $T(p\prime)$
both the drag and diffusion co-efficients can be evaluated from
the following expression:
\begin{eqnarray}
<<T(p)>>=\frac{1}{512\pi^4×} \frac{1}{E_p} \int_{0}^{\infty} \frac{q^2 dq d(cos\chi)}{E_q} \hat{f}(q) \nonumber \\
\frac{\lambda^{\frac{1}{2}}(s,m_p^2,m_q^2)}{\sqrt{s}} \int_{1}^{-1} d(cos\theta_{c.m.})  \nonumber \\ 
\frac{1}{g} \sum  \overline{|M|^2} \int_{0}^{2\pi} d\phi_{c.m.} T(p\prime)
\label{transport}
\end{eqnarray}
where $\lambda(x,y,z)=x^2+y^2+z^2-2xy-2yz-2zx$ 
is the triangular function and $\overline{\mid M\mid^2}$  
in the present case corresponds
to the scattering of $D$-mesons from the light mesons $\pi$, $K$, $\eta$
and nucleons. In a hot pion gas this was obtained  
by Fuchs {\it et al.} 
~\cite{cf} using experimental information on $D$-meson resonances.
Here we evaluate these amplitudes using a covariant 
formulation of chiral perturbation theory in which the leading term
allows for $D$ scattering via $D^*$ meson exchanges in addition to
a contact interaction.  In nuclear matter the $DN$ scattering amplitudes
have been obtained in a coupled channel Bethe-Salpeter approach
where $\Lambda_c$ and $\Sigma_c$ appear as dynamically generated~\cite{tolos}. 
In this work we have obtained  the $DN$ scattering amplitudes proceeding 
via  $\Lambda_c$ and $\Sigma_c$ exchanges using the Lagrangian of Ref.~\cite{Liu_plb}.
The generic Feynman diagrams 
for the elastic processes are depicted in Fig.~\ref{fig0}.
We have included form factors in each of the interaction vertices 
to take into account the finite size of the hadrons. For the $t$ and 
$s$-channel diagrams the form factors are taken as~\cite{Liu_plb}
$F_t=\Lambda^2/(\Lambda^2+\vec{q}^2)$  and
$F_s=\Lambda^2/(\Lambda^2+\vec{p_i}^2)$ respectively, where
$\vec{q}$ is the three momentum transfer and $p_i$ is
the initial three momentum of the light mesons (pion, kaon and eta) 
or nucleon. In the four point (contact) vertices a form factor,
$F_4=\left(\Lambda^2/(\Lambda^2+\bar{q}^2)\right)^2$ with $\bar{q}=
5p_i^2/3$ has been introduced~\cite{liu_npa}. We have taken $\Lambda=1$ GeV.  
The interaction Lagrangian as well as the  $\overline{\mid M\mid^2}$ 
for various processes are detailed in the appendix.

\begin{figure}[ht]
\begin{center}
\includegraphics[scale=0.35]{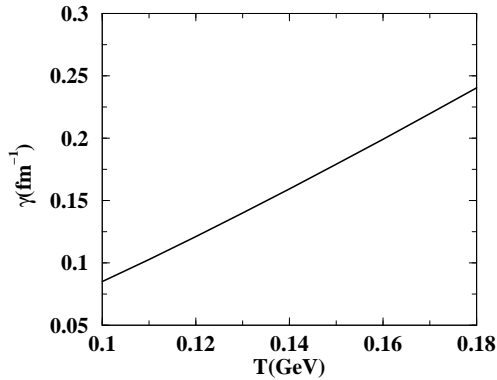}
\caption{The variation of drag coefficients with temperature
due to the interaction of the 
$D$ with thermal pions, nucleons, kaons and eta. 
}
\label{fig1}
\end{center}
\end{figure}

In  Fig.~\ref{fig1} variation of the drag coefficient with temperature 
has been depicted for $D$-mesons. 
We have observed that the $D-\pi$ meson interaction plays 
the most dominant role
in the drag coefficient primarily because of the larger
phase space density of the pions. However,
at higher $T$ the contribution from 
the nucleons become significant.
As mentioned before, $\gamma$ is the thermal average of the
square of the invariant amplitude weighted by
the momentum transfer. Therefore, as the temperature of the thermal bath
increases the hadrons  move faster
and gain the ability to transfer larger momentum during their
interaction with the $D$ mesons - resulting in
the increase of the drag coefficient.
This trend is clearly observed in Fig.~\ref{fig1}.
It may be mentioned here that the drag increases with
$T$ when the system
behaves like a gas. In case of a liquid the
drag may decrease with temperature (except for very few cases) 
since a substantial part of the thermal
energy goes into making the attraction between the interacting
particles weaker. This allows them to move more freely resulting in
a smaller drag force. Therefore, the variation of the drag 
with $T$ may be used to characterize the nature of 
interaction of the fluid. 
The large value of the drag coefficient indicates
that the interaction of the $D$ meson with the thermal medium
is quite significant so that the 
the $D$ meson may get thermalized
in the system and flows with the bulk matter. 
This may be examined by analyzing the 
transverse momentum spectra of the $D$ mesons produced in
heavy ion collisions~\cite{starD}. We find that the typical value of the
relaxation time ($\sim \gamma^{-1}$) is about 4-5 fm/c. Therefore, 
if the life time of the hadronic phase is more than this time 
scale then the $D$ meson may get thermalized indeed.
\begin{figure}[ht]
\begin{center}
\includegraphics[scale=0.35]{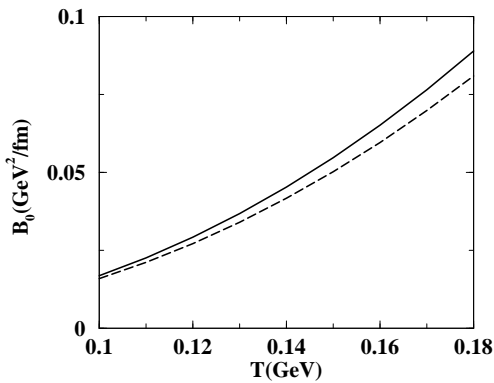}
\caption{Variation of diffusion co-efficient as a function of temperature.
The solid line indicates the variation 
of the diffusion coefficient with temperature
obtained from Eqs.~\ref{diffusion} and~\ref{transport}.
The dashed line stands for the diffusion coefficient
obtained from the Einstein relation (Eq.~\ref{Einstein}).
}
\label{fig2}
\end{center}
\end{figure}


In the same way it may
be argued that the diffusion coefficient involves the square of
the momentum transfer - which should also increase with
$T$ as seen in Fig.~\ref{fig2}. 
The dominant contribution comes from the interaction
of $D$ mesons with pions. 
The drag  and the diffusion coefficients are related 
through the Einstein  relation as: 
\be
B_0=M_D\gamma T.
\label{Einstein}
\ee
where $M_D$ is the mass of the $D$-meson. 
The temperature variation of the diffusion coefficient obtained from 
Eq.~\ref{Einstein}  is depicted in Fig.~\ref{fig2} (dashed line). The  
difference between the results obtained from Eq.~\ref{transport} 
and Einstein's relation is about $10\%$
at $T=180$ MeV. This small difference 
illustrates the validity of the Einstein relation
in the current situation. The value of the spatial diffusion
coefficient, $D_x$ may be expressed 
in terms of drag coefficient as $D_x=T/(M_D\gamma)$. The value of 
$D_x$ at $T=180$ MeV 
is $\sim 2/(2\pi T)$ {\it i.e.} 2 times larger than the thermal
wave length, $\lambda=1/(2\pi T)$ which is well within the quantum bound.

The magnitude of the energy dissipation of the $D$ meson 
in the system may be estimated by using the relation
\be
-\frac{dE}{dx×}=\gamma p~.
\label{dedx}
\ee
The magnitude of $\gamma$ obtained in the present calculation  
indicates a substantial loss of energy of the
$D$ meson in the medium, which might have observable effects on
quantities such as the nuclear suppression factor of single
electrons originating from the decays of heavy mesons.

Here it is necessary to point out that
though chiral perturbation theory provides a consistent framework for 
performing
perturbative calculations of strong interaction processes such as
$D-h$ scattering in this case, it
is limited by the abundance of coupling constants appearing
in the Lagrangian which have to be determined from experimental
data~\cite{Casalbuoni}. In the
case at hand,  the experimental error in the $D^*$ decay width leads to 
an uncertainty ($11\%$) in the $D^*D\pi$ coupling, $g$~\cite{Geng},
which results in a significant variation in the value of the drag
diffusion coefficients.
Interestingly, the lower
bound in this value leads to a drag coefficient which agrees reasonably 
well at high temperatures with that obtained by He et al~\cite{MinHe} using empirical
elastic scattering amplitudes.
In addition to this, chiral symmetry breaking
effects in the pseudoscalar decay constants could also contribute to the
uncertainty. However, since in our case the pion contribution dominates maximally,
this effect will be insignificant.

To summarize, in this work we have evaluated the drag and diffusion coefficients
of open charm mesons propagating in a hadronic background composed of 
pions, kaons, nucleons and eta. We observe that the values of both the 
transport coefficients increases with temperature and the dominant contributions
come from the pions in the medium. However, at higher $T$
the contributions from heavier hadrons become significant.
The magnitude of the drag coefficient
of the $D$ meson in the hadronic medium reveals
that while evaluating the nuclear suppression  for the single
leptons originating from the decays of $D$ mesons the hadronic
contributions should be included.
Lattice QCD calculations~\cite{bazavov} indicate that at
low baryonic chemical potential and high temperature domain
there is no phase transition 
between hadronic matter and QGP - it is a cross over, which means 
that the hadronic matter can make a continuous transition 
to QGP in this region of phase diagram. Therefore, the transport
coefficients evaluated for the hadronic matter with zero baryonic
chemical potential may have vital effects from the quark gluon plasma.   

\section{Appendix}

In this appendix we provide the interaction Lagrangian and matrix elements for
scattering of $D$ mesons from the light mesons ($\pi,K,\eta$) and nucleons discussed in this work.

The leading order chiral Lagrangian describing the interaction of Goldstone
bosons with the heavy-light pseudoscalar ($P$) and vector ($P_\mu^*$) mesons
is given by \cite{Geng}
\bea
{\cal L}_{PP^*\Phi}&=& \langle{\D_\mu P \D^\mu P^\dagger}\rangle -
m_D^2\langle{P P^\dagger }\rangle \nonumber\\
&&- \langle{ \D_\mu P^{*\nu} \D^\mu P^{*\dagger}_\nu }\rangle
+ m_{D^*}^2\langle{P^{*\nu} P^{*\dagger}_\nu }\rangle  \nonumber\\
&&+ ig\langle{P^{*}_{\mu} u^\mu P^{\dagger}-P u^\mu P^{*\dagger}_\mu }\rangle   
\label{eq:Lagtot}
\eea
where $P=(D^{0}, D^{+},D^{+}_s)$ and $P_\mu^*=(D^{*0}_\mu, D^{*+}_\mu,
D^{*+}_{s\mu})$ and $\langle...\rangle$ denotes trace in flavour space.
The covariant derivatives are defined as
$\D_\mu P_a = \del_\mu P_a - P_b\Gm^{ba}_\mu$ and
$\D^\mu P_a^\dag = \del^\mu P_a^\dag + \Gm_{ab}^\mu P_b^\dag$
with $a,b$ the $SU(3)$ flavour indices.
The value of the heavy-light pseudoscalar-vector
coupling constant $g=1177\pm 137$ MeV is obtained by reproducing the
experimental $D^*\rightarrow D\pi$ decay width of $\sim$ 65$\pm 15$ keV
with the above interaction.
The vector and axial-vector currents are respectively
given by $\Gm_\mu=\frac{1}{2}( u^\dagger\del_\mu u + u\del_\mu u^\dagger)$
and $u_\mu=i( u^\dagger\del_\mu u - u\del_\mu u^\dagger)$
where $ u = \exp(\frac{i\Phi}{2\F})$. The unitary matrix $\Phi$ collects the
Goldstone boson fields and is given by
$\Phi=\sqrt{2}\left(\begin{array}{ccc}
\frac{\pi^0}{\sqrt{2}}+\frac{\eta}{\sqrt{6}}
& \pi^{-} & K^{-} \\ \pi^{+}
& -\frac{\pi^0}{\sqrt{2}}+\frac{\eta}{\sqrt{6}}
& K^0 \\ K^+ & K^0 & -\frac{2\eta}{\sqrt{6}}\end{array}
\right)$. To lowest order in $\Phi$ the vector and axial-vector currents are
\be
\Gm_\mu=\frac{1}{8\F^2} [\Phi , \del_\mu \Phi], ~~~~~u_\mu=-\frac{1}{\F}
\del_\mu \Phi~.
\ee

The above interaction allows elastic scattering of the $D$ meson
with the $\pi,K$ and $\eta$ fields through a heavy-light vector meson exchange
in addition to a contact interaction as shown in Fig.~(\ref{fig0}). The form of the contact interaction
obtained in the covariant formulation of chiral perturbation
theory used here coincides with that in~\cite{Lutz}. The invariant amplitudes
for $D^+$ elastic scattering from hadrons, $h$
($D^+(p_1)+h(p_2)\rightarrow D^+(p_3)+h(p_4)$) 
have been obtained as follows:

\bea
&&\overline{|{M}_{D^+\pi^+}|^2}=[\frac{2g^2}{\F^2}
\frac{\{p_1\cdotp p_4 - (p_1\cdotp p_3-m^2_{\pi})^2/m^2_{D^{*}}\}}{t-m^2_{D^{*}}}
\nonumber\\
&&-\frac{1}{4\F^2}(s-u)]^2 
\eea

\bea
&&\overline{|{M}_{D^+\pi^-}|^2}=[\frac{2g^2}{\F^2}
\frac{\{p_1\cdotp p_3 - (p_1\cdotp p_2+m^2_{\pi})^2/m^2_{D^{*}}\}}{s-m^2_{D^{*}}}
\nonumber\\
&&+\frac{1}{4\F^2}(s-u)]^2
\eea

\bea
&&\overline{|{M}_{D^+\pi^0}|^2}=\left(\frac{g^2}{\F^2}\right)^2[
\frac{\{p_1\cdotp p_3 - (p_1\cdotp p_2+m^2_{\pi})^2/m^2_{D^{*}}\}}{s-m^2_{D^{*}}}
\nonumber\\
&&+\frac{\{p_1\cdotp p_4 - (p_1\cdotp p_3-m^2_{\pi})^2/m^2_{D^{*}}\}}{t-m^2_{D^{*}}}]^2
\eea

\bea
&&\overline{|{M}_{D^+\eta}|^2}=\left(\frac{g^2}{3\F^2}\right)^2[
\frac{\{p_1\cdotp p_3 - (p_1\cdotp p_2+m^2_{\eta})^2/m^2_{D^{*}}\}}{s-m^2_{D^{*}}}
\nonumber\\
&&+\frac{\{p_1\cdotp p_4 - (p_1\cdotp p_3-m^2_{\eta})^2/m^2_{D^{*}}\}}{t-m^2_{D^{*}}}]^2
\eea

\bea
&&\overline{|{M}_{D^+K^0}|^2}=[\frac{2g^2}{\F^2}
\frac{\{p_1\cdotp p_3 - (p_1\cdotp p_2+m^2_{K^0})^2/m^2_{D^{*}_s}\}}{s-m^2_{D^{*}_s}}
\nonumber\\
&&+\frac{1}{4\F^2}(s-u)]^2 
\eea

\bea
&&\overline{|{M}_{D^+\oK^0}|^2}=[\frac{2g^2}{\F^2}
\frac{\{p_1\cdotp p_4 - (p_1\cdotp p_3-m^2_{\oK^0})^2/m^2_{D^{*}_s}\}}{t-m^2_{D^{*}_s}}
\nonumber\\
&&-\frac{1}{4\F^2}(s-u)]^2 
\eea

In the numerical calculations we have used the physical masses for the particles
involved. For the heavy-light mesons we have taken $m_D=1867$ MeV, $m_{D^*}=2008$
MeV and $m_{D_s}=1969$ MeV.

We next discuss scattering of the $D$ meson with nucleons
which proceeds via exchange of the charmed baryons $\Sigma_c$
and $\Lambda_c $ according to the Lagrangian~\cite{Liu_plb},
\bea
{\cal L}_{DN}&=&\frac{f_{DN\Lambda_c}}{m_D}[\overline{N}\gamma^5\gamma^\mu
\Lambda_c\del_\mu \overline D +
\del_\mu D\bar{\Lambda}_c \gamma^5\gamma^\mu N] \nonumber\\
&&+\frac{f_{DN\Sigma_c}}{m_D}[\overline{N}\gamma^5\gamma^\mu\,(\vec\tau\cdot
\vec\Sigma_c)\,\del_\mu \overline D
\nonumber\\
&&+ \del_\mu D\,(\vec\tau\cdot\vec{\overline\Sigma_c})\, \gamma^5\gamma^{\mu} N] 
\eea
where $D=(D^0,D^+)$ and $\overline{D}=(\overline{D^0},D^-)^T$.
The amplitudes for $D^+$ mesons elastically scattering from nucleons (and anti-nucleons)
via $\Lambda_c$ exchange are obtained as,

\be
\overline{|{M}_{D^+n}|^2}=\frac{1}{2}
(\frac{f_{DN\Lambda_c}}{m_D})^4 X^{s}_{DN\Lambda_c}(m_{D},m_N,m_{\Lambda_c}) \nonumber\\
\ee
\be
\overline{|{M}_{D^+\bar n}|^2}=\frac{1}{2}
(\frac{f_{DN\Lambda_c}}{m_D})^4 X^{t}_{DN\Lambda_c}(m_{D},m_N,m_{\Lambda_c}) \nonumber\\
\ee
and those proceeding via $\Sigma_c$  are as
\be
\overline{|{M}_{D^+p}|^2}=\frac{1}{2}
(\frac{\sqrt{2}f_{DN\Sigma_c}}{m_D})^4 X^{s}_{DN\Sigma_c}(m_{D},m_N,m_{\Sigma_c})
\ee
\be
\overline{|{M}_{D^+n}|^2}=\frac{1}{2}
(\frac{f_{DN\Sigma_c}}{m_D})^4 X^{s}_{DN\Sigma_c}(m_{D},m_N,m_{\Sigma_c})
\ee
\be
\overline{|{M}_{D^+\bar p}|^2}=\frac{1}{2}
(\frac{\sqrt{2}f_{DN\Sigma_c}}{m_D})^4 X^{t}_{DN\Sigma_c}(m_{D},m_N,m_{\Sigma_c})
\ee
\be
\overline{|{M}_{D^+\bar n}|^2}=\frac{1}{2}
(\frac{f_{DN\Sigma_c}}{m_D})^4 X^{t}_{DN\Sigma_c}(m_{D},m_N,m_{\Sigma_c})
\ee

where
\bea
&& X^{s}_{DNB}(m_{D},m_N,m_B)=\frac{4}{(s-m^2_B)^2}[m_B^2\{m_N^4m_{D}^2
\nonumber\\
&&-2(p_1\cdotp p_2)(p_1\cdotp p_4)(m^2_N+m_{D}^2)+(p_1\cdotp p_3)(4(p_1\cdotp p_2)^2
\nonumber\\
&& +m_{D}^2 m^2_N)\} ~ + ~ m_B\{2(m_N^3m_{D}^2-2(p_1\cdotp p_2)^2m_N)
\nonumber\\
&&~~~~~~(m_N^2+m_{D}^2) -4m_Nm_D^4(p_1\cdotp p_2)\}
\nonumber\\
&& + \{8(p_1\cdotp p_2)((p_1\cdotp p_2)+m_{D}^2)((p_1\cdotp p_2)^2-m_{D}^2m_N^2)
\nonumber\\
&& +2m_{D}^4((p_1\cdotp p_2)+m_{D}^2)^2\} ~ + ~ s\{m_N^4m_{D}^2 + 4 m_{D}^2
\nonumber\\
&& (p_1\cdotp p_2)(p_1\cdotp p_4) - (m_{D}^4 + 4 (p_1\cdotp p_2)^2)(p_1\cdotp p_3)\}]
\nonumber\\
\eea
and
\bea
&& X^{t}_{DNB}(m_{D},m_N,m_B)=\frac{4}{(t-m^2_B)^2}[m_B^2\{m_N^4m_{D}^2
\nonumber\\
&&-2(p_1\cdotp p_3)(p_1\cdotp p_2)(m^2_N+m_{D}^2)+(p_1\cdotp p_4)(4(p_1\cdotp p_3)^2
\nonumber\\
&& +m_{D}^2 m^2_N)\} ~ + ~ m_B\{2(-m_N^3m_{D}^2+2(p_1\cdotp p_3)^2m_N)
\nonumber\\
&&~~~~~~(m_N^2+m_{D}^2) +4m_Nm_D^4(p_1\cdotp p_3)\}
\nonumber\\
&& + \{8(p_1\cdotp p_3)((p_1\cdotp p_3)+m_{D}^2)((p_1\cdotp p_3)^2-m_{D}^2m_N^2)
\nonumber\\
&& +2m_{D}^4((p_1\cdotp p_3)+m_{D}^2)^2\} ~ + ~ t\{m_N^4m_{D}^2 + 4 m_{D}^2
\nonumber\\
&& (p_1\cdotp p_3)(p_1\cdotp p_2) - (m_{D}^4 + 4 (p_1\cdotp p_3)^2)(p_1\cdotp p_4)\}]
\nonumber\\
\eea
The coupling constants obtained using $SU(4)$ symmetry
are given by~\cite{Liu_plb}  $\frac{f_{DN\Lambda_c}}{m_D}=7.18$ GeV$^{-1}$ and
$\frac{f_{DN\Sigma_c}}{m_D}=2.01$ GeV$^{-1}$.

The various invariant amplitudes can be expressed in terms of 
the Mandelstam variables using the relations:
$p_1 \cdot p_2 =\frac{s-m_{D}^2 -m_h^2}{2} $,  $p_1 \cdot p_3 =\frac{m_{D}^2 + m_h^2 -t}{2} $  and
$p_1 \cdot p_4 =\frac{m_{D}^2 + m_h^2 -u}{2} $ 
where $m_D$ is the mass of $D$ meson and $m_h$ that of the light hadrons.

{\bf Acknowledgment:} 
We thank Ralf Rapp for very useful and pertinent comments.
SKD and JA partially  supported by DAE-BRNS project Sanction No.  2005/21/5-BRNS/2455.

\end{document}